\documentclass[fleqn,12pt]{wlscirep}
\usepackage{wrapfig, framed, caption}
\usepackage[utf8]{inputenc}
\usepackage{heuristica}
\usepackage{graphicx}
\usepackage{wrapfig, framed, caption}
\usepackage{tcolorbox}
\usepackage{xr}
\usepackage{hyperref}
\usepackage{xcolor}

\makeatletter
\newcommand*{\addFileDependency}[1]{% argument=file name and extension
  \typeout{(#1)}
  \@addtofilelist{#1}
  \IfFileExists{#1}{}{\typeout{No file #1.}}
}
\makeatother

\usepackage{setspace}
\usepackage[normalem]{ulem}
\useunder{\uline}{\ulined}{}
\usepackage{xparse}
\newsavebox{\fminipagebox}
\NewDocumentEnvironment{fminipage}{m O{\fboxsep}}
 {\par\kern#2\noindent\begin{lrbox}{\fminipagebox}
  \begin{minipage}{#1}\ignorespaces
 \end{minipage}\end{lrbox}%
  \makebox[#1]{%
    \kern\dimexpr-\fboxsep-\fboxrule\relax
    \fbox{\usebox{\fminipagebox}}%
    \kern\dimexpr-\fboxsep-\fboxrule\relax
  }\par\kern#2
 }

\title{Network Science Predicts Who Dies Next in Game of Thrones}

\author[*1,2,3]{Mil\'an Janosov}

\affil[1]{Department of Network and Data Science, Central European University, Budapest, 1051, Hungary}

\affil[2]{Datapolis Inc, Budapest, 1112, Hungary}

\affil[3]{Milan Janosov \href{https://linktr.ee/floraborsi}{https://linktr.ee/janosov}}

\affil[*]{janosovm@gmail.com}

\begin{document}

\maketitle

%%% ============================================================================================================
%%% -------------------------------------------------     INTRO
%%% ============================================================================================================

\section*{Abstract}
{\small

Social network analysis and machine learning have found countless applications in recent years. As an example, this short project was carried out in 2017 and was followed by significant media attention, with the following goal: to bring network science and predictive modeling together on the subject of the popular TV and book series, Game of Thrones, and predict which key characters are likely to meet their ends.

}

\vspace{0.5cm}
{\small {\bf Keywords}: Game of Thrones, network science, social network analysis, binary prediction}

\section{Introduction}

This article summarizes two blogposts published in 2017~\cite{post1,post2} that aimed to combine network science~\cite{netsci}, predictive modeling, and the TV show (and book series adaptation) Game of Thrones to pinpoint the major characters that are likely to meet their ends in the then-coming episodes of the series. Due to the high interest received by various media outlets ~\cite{feat1,feat2,feat3,feat4,feat5,feat6,feat7}, the author here summarizes this project. 

In this project, I built on the labeled subtitles of the TV series, which included not only the spoken texts but also the names of the individuals speaking. Additionally, the dataset used~\cite{data} had information on the separators between the different scenes. By combining these two pieces of information, I extracted the list of characters appearing in every scene - altogether about 600. Then I assumed that these scenes are the basic building blocks of the show's social system and used them to construct the social map of Westeros as follows. In this network, every major character is represented by a network node, while there is a link between two players if they co-occurred in the same scene. Additionally, the more frequently they co-occurred, the stronger their connection. Then I computed the values of different network centrality measures of these characters (network nodes) to use as prediction features and labeled the characters whether they died or not in the first six seasons. Finally, I applied a widely used linear model, support vector machine, to predict which of the still-living characters were likely to meet their ends.

The second section details the original prediction built on the subtitles of the first six seasons, while the third section updates the prediction after season 7. The summary, added later, provides a brief evaluation of the final results.

\section{Game of Thrones Prediction}

The new season of Game of Thrones is almost upon us and fans are excited about what it may bring. I am probably not alone in wondering which of my favourite characters are going to meet their ends, and which will live on to the next season. So I decided to come up with a ranking for the characters based on how likely it is that they will die. Game of Thrones is a complex world in which social position and true friends seem to be quite important, so I quantified each character’s social interaction patterns using the tools of network science. I then predict their fate using machine learning methods.

\subsection{Creating the network of Westeros}

As a data source I used the show’s subtitles, collected in dialogue format on a fan website~\cite{data}. Unfortunately most of the episodes from season two and three are missing but the remaining four seasons, including almost 600 scenes, are available in a consistent format.

\begin{figure}[!hbt]
\centering
\includegraphics[width=1.00\textwidth]{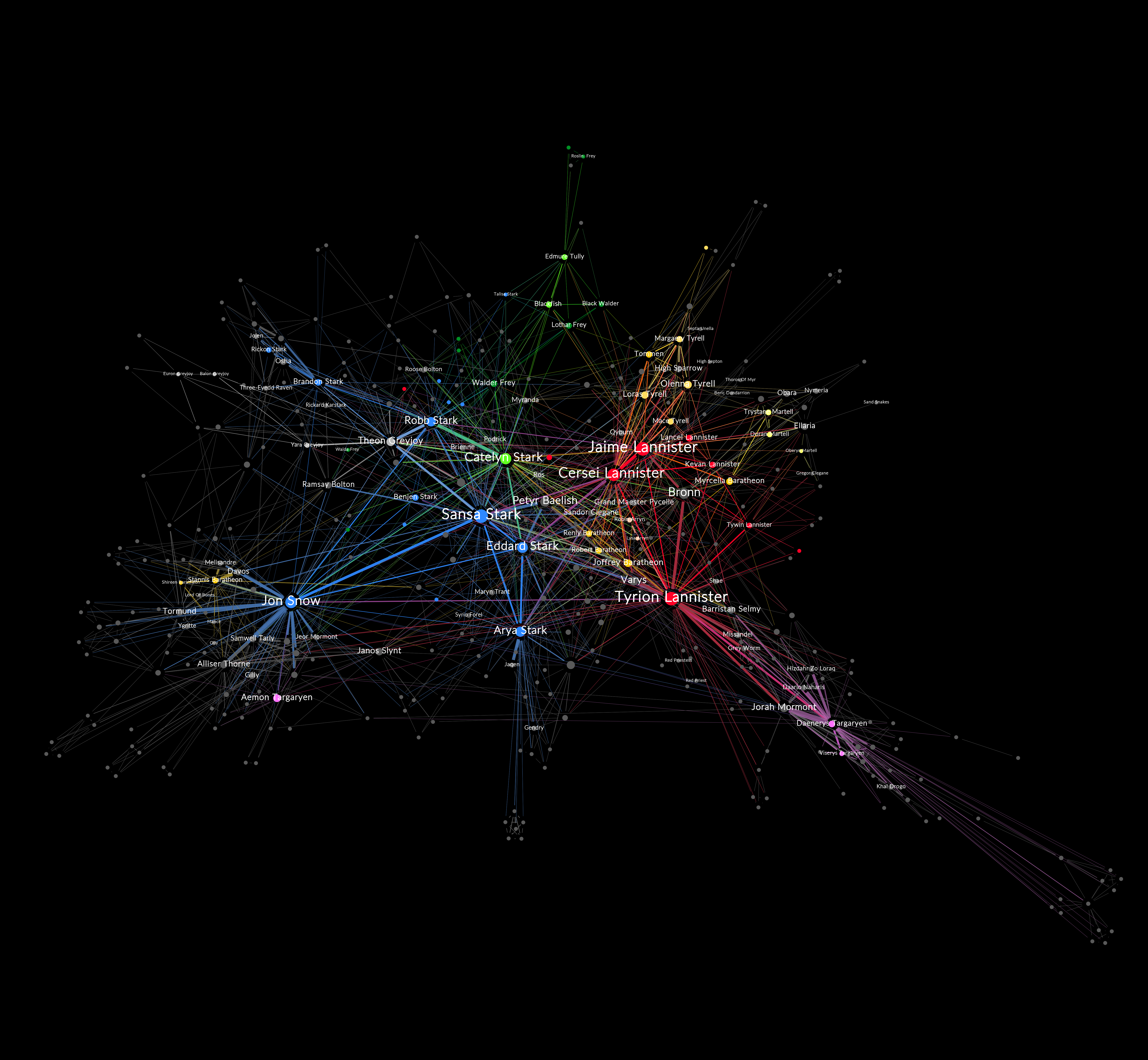}
\caption{The overall network map of Westeros based on the first six seasons of the Game of Thrones TV show.}
\label{fig:fig1}
\end{figure}

First I constructed the aggregated network of the realm’s social system. In this network each node represents a character of the story, and the weight of the link between each pair of characters symbolizes the strength of their social interaction. I considered scenes to be the elementary units of the social interaction (an average episode contains about twenty of them). This means that everyone who appeared once (twice) together in the same scene has a tie with strength of one (two), and within a scene everyone is connected with everyone. In other words, scenes are complete graphs, or cliques, increasing the tie strength between all pairs of people present by one. By calculating these  scene-level complete networks and then aggregating them, we arrive to the global social network of Westeros (Figure \ref{fig:fig1}), which has almost 400 nodes and more than 3000 edges.

In the network visualization (Figure \ref{fig:fig1}) all the members of the great houses are marked with different colors (e.g. blue – Starks, red – Lannisters, yellow – Martells), while the rest of the people are in gray.  The size of the nodes is proportional to the number of contacts each person has and the names of the most popular characters are added as labels. The less interesting nodes with a very low degree centrality are filtered out. We can see a separated community around Jon Snow, indicating that the folks around the Wall have only a few contacts with the rest of the realm. Tyrion has a separate role: he connects Daenerys Targaryen to the center of the network, including King’s Landing, where we can see two large communities. These are the Starks and the Lannisters and their zones of influence and interaction, like the bonds between the Stark and Tully families and the conflict between the Lannisters and the Martells, forming a dense web at the heart of the story.

\begin{figure}[!hbt]
\centering
\includegraphics[width=1.00\textwidth]{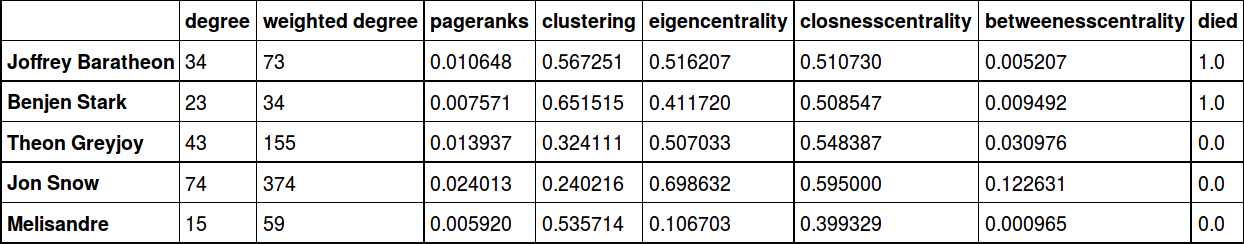}
\caption{The set of features (based on network analysis) and the target variable (whether a certain character died over the first six seasons or not) for six randomly selected characters.}
\label{fig:fig2}
\end{figure}

Let’s turn to the math. We can calculate various measures of how important nodes are. We associate these measurements to the characters to illustrate their importance in this social ecosystem. Some of these measures are i) the node degree - the number of contacts a person has; ii)  the weighted degree - the sum of edge weights at a certain node; iii) the clustering - how often pairs of contacts of the node are in contact themselves ; and iv) the betweenness centrality, which says how much of a bridge a node is in terms of information flow by measuring how often it lies on the shortest path between other pairs of nodes (Figure \ref{fig:fig2}). Besides  getting a better idea of who is important and who is not, we can also learn from the data which characters died in the first six seasons. Our goal then is to relate network position to survival: does one predict the other? In other words, we want to train an algorithm to figure out which network measures predict whether a character has died.

\subsection{Prediction}

We have a set of 94 characters interesting enough to care about. All of them are described by seven different network-based features which proxy for different dimensions of their social importance. We also know which of the characters have already died (61 of them). Based on this knowledge we can form an educated guess of who is going to die in the near future in the following way: which of those people still alive have similar features to those who have already passed away? This problem resembles the so-called churn problem (widely studied in data science), which can be solved with various classification-based algorithms. In this analysis we use a Support Vector Machine (SVM), which happened to be the most accurate. It has an easy-to-use implementation in Python in case you’d like to try this at home~\cite{sk}.

The machine learning algorithm takes all features into account and makes predictions on the possible value of the target variable. For this the sample data is split into test and training sets randomly and multiple times, the prediction is made on all the random splits, and the final result is evaluated. With this cross-validation strategy the SVM classifier predicted the correct class (dead or alive) in 72.3\% of the cases, which given the size and nature of the data is a fair result. To illustrate the accuracy, the model says that eight characters shouldn’t have died but in the story they did - the model couldn’t foresee their death. Such characters are e.g. Margaery Tyrell - death of queens seems to be less likely than that of kings,  and Janos Slynt who was exiled from Kings Landing to the Wall, where his powerful friends couldn't save him, even though the model suggested so.

It should be mentioned that including other types of features (e.g. gender, being a member of a noble house, or sentiment analysis of the speeches), having a more complete dataset, comparing the TV show to the book, etc., could increase the accuracy of the predictions. This model also neglects discrepancies like Jon Snow dying and then being reborn, and Benjen Stark being somewhere in between.

\subsection{Results – Spoiler alert}

Using the SVM model we  get to the answers – the probabilities of each living well-known character passing away. As network measures are often very correlated, we can’t pick one or two that are highly predictive on their own, but seemingly characters with high betweenness, low clustering and high degree are less likely to be killed. In any case the strength of the machine learning approach is exactly finding hidden relationships among the large number of features. I used five-fold cross-validation during the prediction, and repeated this a hundred times to get an estimation on the statistical value and error of the probabilities. Finally, here is the list of characters ranked in increasing order of survival according to the final prediction model (Figure \ref{fig:fig3}):

\begin{figure}[!hbt]
\centering
\includegraphics[width=0.30\textwidth]{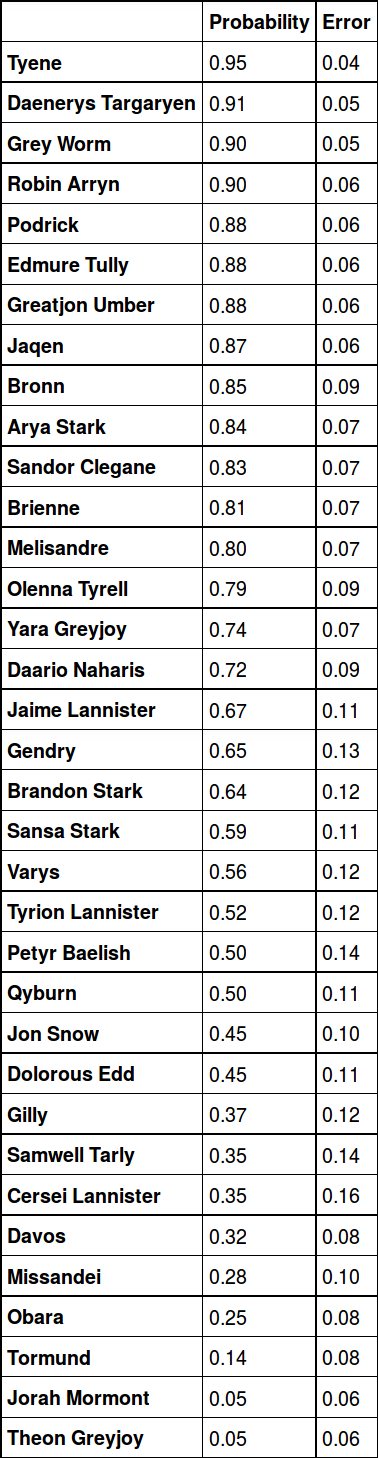}
\caption{Game of Thrones characters and their probability of dying based on their network centrality patterns, and the error of these probabilities based on repeating the prediction (with five-fold cross-validation) a hundred times.}
\label{fig:fig3}
\end{figure}

The list tells us many interesting things. First, Daenerys seems to be quite likely to die, overlapping with a number of speculations, while Tyrion and Jon Snow seem to be relatively safe. Second, both the ever-popular Arya Stark and the less friendly Hound, already so close to death many times before, are both in dangerous positions. Surprisingly, Cersei, currently sitting on the Iron Throne, and Baelish who is doing his best to get there, seem to be in a much better position. It seems Jorah Mormont will find the cure for his greyscale disease, and despite all he has been through, Theon Greyjoy will probably survive. Sadly, the same cannot be said about the Arryn family.

\clearpage
\newpage

\section{Game of Thrones Prediction 2. 0}

The 7th season of Game of Thrones, which brought much unexpected news and started countless debates – and naturally killed some important characters – is over. So here is the first chance to validate my previous prediction on who would possibly die in the show.

\subsection{What has happened so far?}

In the former blog post I used the TV show's episodes to create a social network of the realm of Westeros, determined various centrality parameters of both the dead and the living characters, and used a support vector machine model to predict the likelihood of each living character to die. The result came in the form of a list of characters ranked by their chances to die before the end of the series. My prediction disregards time: it does not foresee whether a character would die right away in the first episode or just at the very end. Thus, even though the final season is still to come, I would like to discuss the preliminary results of the model.
\begin{figure}[!hbt]
\centering
\includegraphics[width=0.35\textwidth]{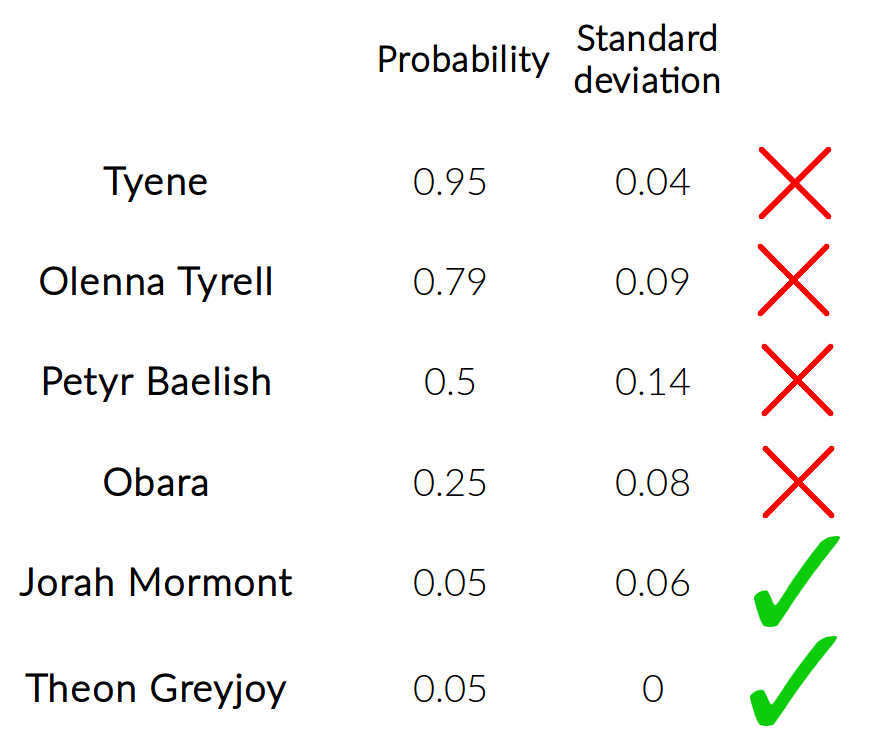}
\caption{The characters are most seriously affected by the new season and their predicted probabilities of dying. The red cross means the character passed away in the new season, while the green tick indicates its survival, showing that characters above 50\% passed away while low-probability individuals survived.}
\label{fig:fig4}
\end{figure}
The previous prediction covered the fate of 34 well-known characters, the majority of whom appeared in the new season as well. However, only four of them have died so far, which, if the deaths had happened randomly, would mean an approximate 11\% of chance of dying for all the characters. Despite the model's limitations and simplicity, it has done a fair job (Figure \ref{fig:fig4}): predicted the death of the first on the list, Tyene, as well as Olenna Tyrell and Petyr Baelish; however, it failed to forecast Tyene’s sister’s death. On the other hand, it was right to assume that Jorah somehow would find the cure for his (thought-to-be) lethal sickness and survive. Likewise, it was correct about Theon: despite all he has been through, he still carries on.

\subsection{Prediction 2.0}

\begin{figure}[!hbt]
\centering
\includegraphics[width=0.75\textwidth]{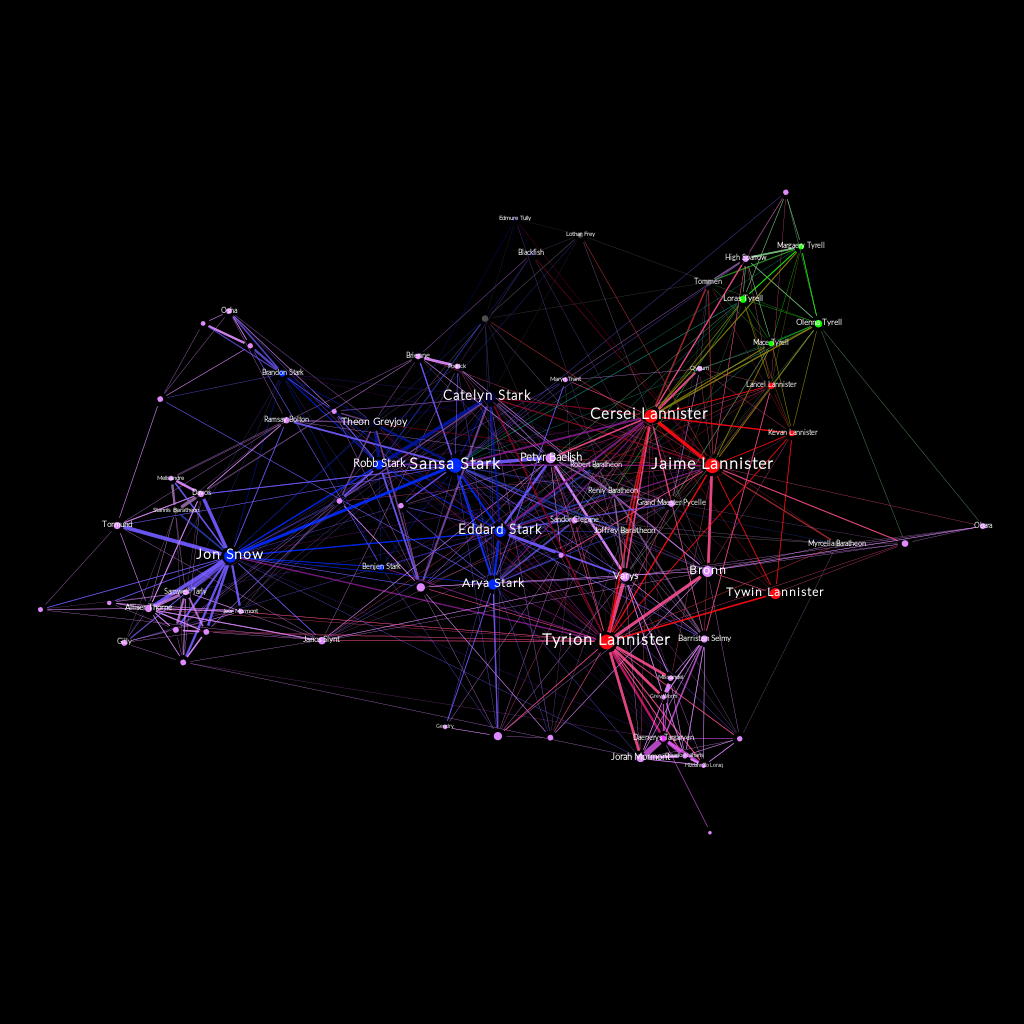}
\caption{The social network of the Game of Thrones realm based on Seasons 1-6, filtering out the less important characters. The size of the nodes is proportional to the strength of the nodes, while the colors are assigned to the great houses.}
\label{fig:fig5}
\end{figure}

Using the previously introduced methodology I also calculated the current ranking of characters based on their probability to die, which as a light spoiler can be found at the end of this post.

\begin{figure}[!hbt]
\centering
\includegraphics[width=0.75\textwidth]{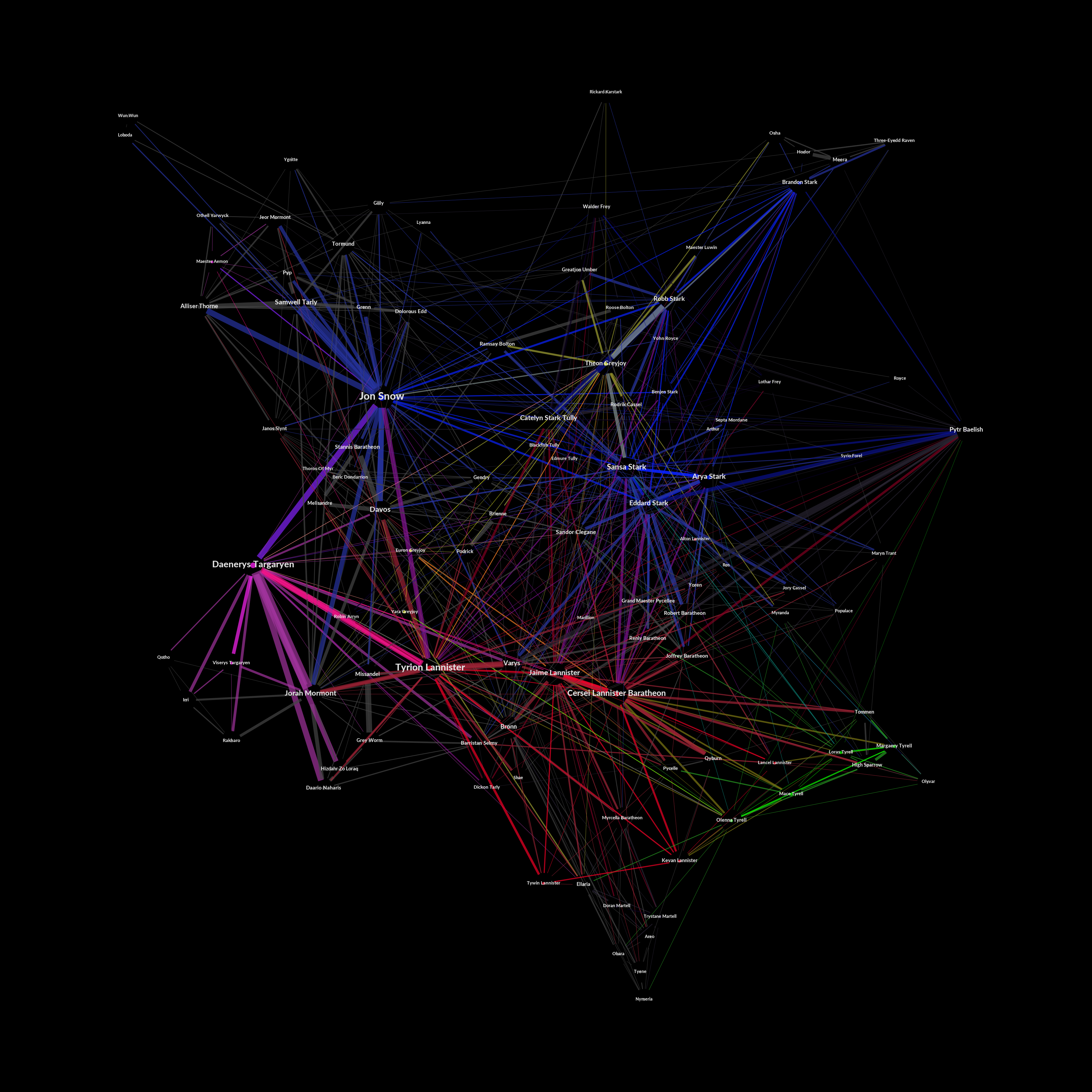}
\caption{Similar to Figure \ref{fig:fig5}, except for Seasons 1-7..}
\label{fig:fig6}
\end{figure}

The most important event is unquestionably Daenerys’s meeting with Jon, and their encounter with Cersei – this is how the previously unconnected domains of this network (Figure \ref{fig:fig5}) finally got connected (Figure \ref{fig:fig6}). This  establishes many new connections and transforms the original social network, which also changes the characters’ predicted future. The filtered network of the central characters based on Seasons 1-6 can be seen in Figure 1, while the network of Seasons 1-7 is in Figure 2. The nodes represent characters with size proportional of the total strength of their connections, and the colors code the great houses appearing in Season 7 (blue – Stark, red – Lannister, magenta – Targaryen, green – Tyrell, yellow – Greyjoy).

\begin{figure}[!hbt]
\centering
\includegraphics[width=0.45\textwidth]{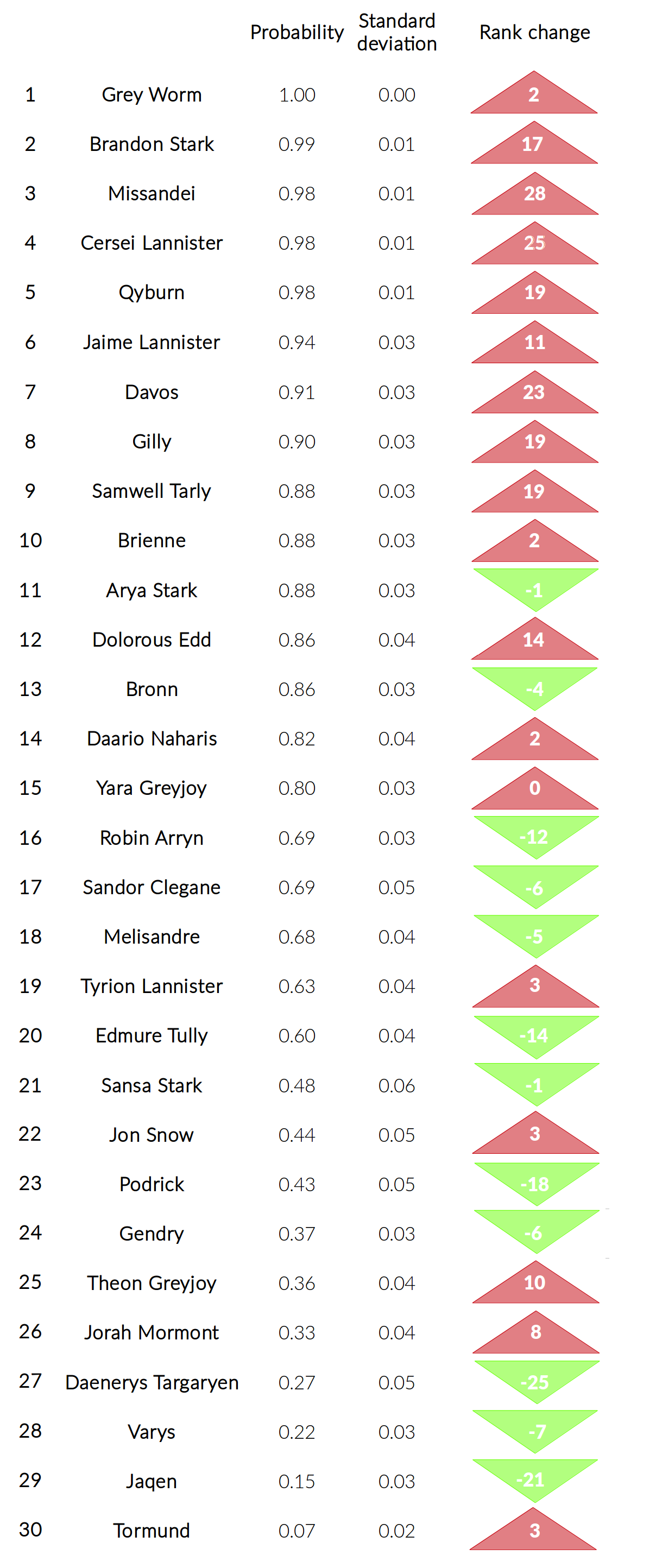}
\caption{The current ranking of the major characters based on their probability to die. The red triangles indicate that their rank moved up on the list (they are more likely to die due to the events of Season 7), while the green triangle means they got into a safer position comparing to the one after Season 6.}
\label{fig:fig7}
\end{figure}

From the list and the network we can deduce several things:
\begin{enumerate}
    \item Jon being the most important character based on almost all network measures;
    \item Bran is finally close to the action which seems to promise a bad future for him;
    \item Daenerys apparently made enough influential friends to make it through; however, those around her might get sacrificed on her mission;
    \item people around Cersei are disappearing (see a much smaller red region, which resembles the Lannisters, in Figure 2), which influences the probabilities for her and those close to her;
    \item Theon and Jorah have a good chance (perhaps based on previous experiences) to survive; and
    \item it seems Tormund, a favorite wilding to many, may not only escape from the falling Wall but survive the entire war.
\end{enumerate}

\section{Conclusion}

\begin{figure}[!hbt]
\centering
\includegraphics[width=0.8\textwidth]{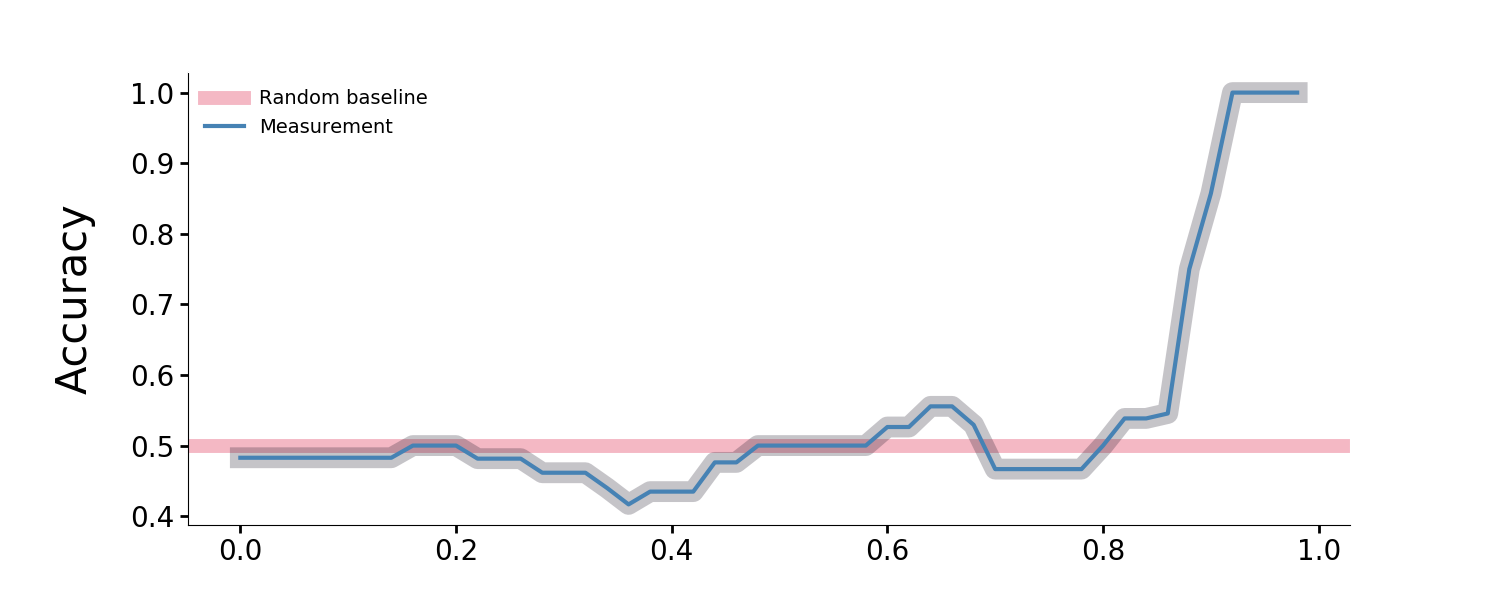}
\caption{The prediction accuracy, as the fraction of the number of correct predictions to the total number of characters with a predicted probability higher than the confidentiality threshold.}
\label{fig:eval}
\end{figure}

From Figure \ref{fig:eval} we can read that the predicted probabilities did not correspond to clear signals below about 87\%  but highly overlap (and fluctuate around) the random baseline (a probability we would get if we were building our prediction on flipping coins). However, we see a quick and clear increase above the threshold of 87\% where the predicted hits highly overlap with the actual ones. This conclusion implies that network features were able to capture the clearest deaths of Game of Thrones. However, it could not work well with uncertain situations. 

The inaccuracies and limitations have several dimensions. First, the cleanness and volume of the data could have been improved. Second, the type of data and features could be extended in the feature, and, e.g., one could enrich the network features by meta-information about the characters (e.g., age, gender) and combine information from the TV series and the original books. Finally, and most interestingly, the predicted period - seasons 7 and 8 - were not based on the book, while the rest of the training data was. This dimension should be very important in terms of how the logic of the show is built up, yet, a very difficult one to study in a quantitative way and certainly a matter of future research.

\section{Data accessibility}
Supporting files associated with this study can be found at \href{https://github.com/milanjanosov/GameOfThrones/}{https://github.com/milanjanosov/GameOfThrones/}

\section{Disclaimer}
While the blog posts are published in sections two and three in their original form, a few minor typos and grammatic issues were corrected without affecting the meaning of the text.

\section{Acknowledgement}
The author wishes to thank Ágnes Diós-Tóth, Thomas Rooney, and Robin Bellers from the Center of Academic Writing at the Central European University for reviewing several parts of this paper and the original blogposts.

\end{document}